\documentclass[conference]{IEEEtran}
\IEEEoverridecommandlockouts
\usepackage{cite}
\usepackage{amsmath,amssymb,amsfonts}
\usepackage{graphicx}
\usepackage{textcomp}
\usepackage{xcolor}
\usepackage{bm}
\usepackage{cuted}
\usepackage{algorithmic, algorithm}
\usepackage{enumitem}
\usepackage{enumitem}
\def\BibTeX{{\rm B\kern-.05em{\sc i\kern-.025em b}\kern-.08em
    T\kern-.1667em\lower.7ex\hbox{E}\kern-.125emX}}

\newcommand*{\herm}{^{\mathsf{H}}}
\newcommand*{\transp}{^{\mathsf{T}}}

\DeclareMathOperator{\diag}{diag}
\DeclareMathOperator{\trace}{Tr}
\DeclareMathOperator*{\argmin}{\arg\min}
\DeclareMathOperator*{\argmax}{\arg\max}

\newcommand{\e}{\mathrm{e}}
\renewcommand{\j}{\mathrm{j}}

\makeatletter
\g@addto@macro\normalsize{%
  \setlength\abovedisplayskip{4.5pt}
  \setlength\belowdisplayskip{4.5pt}
}

\begin{document}

\title{
Leveraging Line-of-Sight Propagation for Near-Field Beamfocusing in Cell-Free Networks
\thanks{The work of G. Mylonopoulos, G. Interdonato and S. Buzzi has been supported by the European Union under the Italian National Recovery and Resilience Plan (NRRP) of NextGenerationEU, partnership on “Telecommunications of the Future” (PE00000001 - program “RESTART”, Structural Project 6GWINET and Spoke 3 Cascade Call Project SPARKS. The work of Pei Liu was supported in part by the National Natural Science Foundation of China under Grant 62471346 and in part by the China Scholarship Council (CSC) under Grant 202306950052.}
}

\author{
\IEEEauthorblockN{Georgios Mylonopoulos$^1$, Giovanni Interdonato$^{1,2}$, Stefano Buzzi$^{1,2,3}$, and Pei Liu$^{4}$}
\IEEEauthorblockA{$^1$Department of Electrical and Information Engineering, University of Cassino and Southern Lazio, 03043 Cassino, Italy.\\
$^2$Consorzio Nazionale Interuniversitario per le Telecomunicazioni (CNIT), 43124 Parma, Italy.\\
$^3$Department of Electronics, Information and Bioengineering, Politecnico di Milano, 20133 Milan, Italy.\\
$^4$School of Information Engineering, Wuhan University of Technology, Wuhan 430070, China.\\
(e-mail: \{georgios.mylonopoulos, giovanni.interdonato, buzzi\}@unicas.it; pei.liu@ieee.org)}
}

\maketitle

\begin{abstract}
Cell-free (CF) massive multiple-input multiple-output (MIMO) is a promising approach for next-generation wireless networks, enabling scalable deployments of multiple small access points (APs) to enhance coverage and service for multiple user equipments (UEs). While most existing research focuses on low-frequency bands with Rayleigh fading models, emerging 5G trends are shifting toward higher frequencies, where geometric channel models and line-of-sight (LoS) propagation become more relevant. In this work, we explore how distributed massive MIMO in the LoS regime can achieve near-field-like conditions by forming artificially large arrays through coordinated AP deployments. We investigate centralized and decentralized CF architectures, leveraging structured channel estimation (SCE) techniques that exploit the line-of-sight properties of geometric channels. Our results demonstrate that dense distributed AP deployments significantly improve system performance w.r.t. the case of a co-located array, even in highly populated UE scenarios, while SCE approaches the performance of perfect CSI.
\end{abstract}


\vspace{-2mm}
\section{Introduction}
\vspace{-2mm}
Cell-free (CF) massive multiple-input multiple-output (m-MIMO) has emerged as a key enabler for next-generation wireless networks, offering scalable, dense deployments of small access points (APs) to enhance coverage and service for multiple user equipments (UEs)~\cite{7827017,ngo2024ultradense}. Various CF architectures have been vastly explored over the past few years, each imposing different requirements on the availability of channel state information (CSI) and on the amount data exchange between APs and central processing units (CPUs)~\cite{bjornson2019making,8901451}. 

CF m-MIMO research has so far mostly focused on low-frequency bands under non line-of-sight (LoS) conditions, i.e.,  employing Rayleigh fading models\cite{8000355,bjornson2019making}. However, as CF networks become denser, LoS conditions are increasingly common. Therefore, there are also recent studies exploring Rician fading models~\cite{10713253,10978201} with dominant LoS links. Moreover, considering dominant LoS links serves to validate performance limits for spatially sparse channels and extreme channel hardening conditions.

Within this framework, geometric channel models become more relevant as they provide a structured representation of the channel based on UE positions relative to APs~\cite{9050553}. A key feature of these models is the distinction between far-field and near-field conditions. As antenna arrays grow larger, individual array components still operate under far-field assumptions, while the entire array transitions to near-field behavior~\cite{10541333,kosasih2024achieving}. In near-field conditions, UEs can be differentiated not only in the angular domain but also by range, increasing system degrees of freedom and improving user capacity~\cite{10615807,9903389}.

In this paper, building on the geometric channel model framework, we introduce a novel perspective on distributed m-MIMO, demonstrating that near-field-like conditions can be achieved through coordinated AP deployments that form artificially large arrays. In this setup, a CF network operates as a distributed m-MIMO system, where the APs serving each UE collectively form a virtual extremely large antenna arrays (V-ELAA) forcing near field channel behavior. Indeed, while individual APs may remain in the far-field, their combined effect enables near-field capabilities, supporting higher user densities. We propose and analyze various channel estimation strategies that leverage the LoS structure of geometric channel models, comparing centralized and decentralized CF architectures. Also, we benchmark these architectures against a conventional m-MIMO BS to assess its performance. Our results show that in the LoS regime, a CF architecture significantly enhances performance, even in highly populated scenarios, and that leveraging the geometric channel structure improves estimation accuracy, approaching the perfect CSI performance.

\textbf{Notation:} Column vectors and matrices are represented by lowercase and uppercase boldface letters, respectively. The operators $(\cdot)\transp$, $(\cdot)^{\ast}$, and $(\cdot)\herm$ denote transpose, conjugate, and conjugate-transpose, respectively. The notation $\mathcal{CN}(\mu,\sigma^2)$ represents a circularly symmetric complex normal distribution with mean $\mu$ and variance $\sigma^2$, while $\mathcal{U}(a,b)$ denotes a uniform distribution over $[a,b]$. Additionally, $\| \cdot \|$ denotes the Euclidean norm, $\j$ is the imaginary unit, and $\mathbb{E}\{\cdot\}$ represents statistical expectation.

\begin{figure}[t]
    \centering
    \includegraphics[width=0.9\columnwidth]{sm_test.png}
    \caption{Considered system architectures, highlighting that multiple APs form a virtual extremely large aperture array (V-ELAA) that preserves the spherical wavefront characteristics of the transmitting UE.}
    \label{fig:sm}
\end{figure}

\section{System Model}
Consider a cell-free system where $M$ APs serve $K$ single-antenna UE devices. Each AP has a uniform planar array (UPA) of $N_{\rm{AP}}$ antennas with inter-antenna spacing $\delta_{o}$. The APs connect to a CPU for data encoding/decoding. Uplink and downlink operate in a time-division duplex (TDD) scheme over a bandwidth $B$ centered at $f_{o}$. The channel coherence spans $T_{\rm{c}}$ intervals, with $T_{\rm{p}}$ for training and $T_{\rm{u}}$ for uplink, where $T_{\rm{c}} = T_{\rm{p}}+T_{\rm{u}}$, disregarding the downlink operations. 

\subsection{Channel Model}
\vspace{-2mm}
Consider the free-space LoS channel between the $k$-th UE and the $n$-th antenna of the $m$-th AP 
\begin{equation}\label{eq:h_knm}
h^{(n)}_{k,m} = \rho\left(d^{(n)}_{k,m}\right)\e^{\j\phi\left(d^{(n)}_{k,m}\right)}\e^{\j\psi_{k}},
\end{equation}
where $\lambda_{o}$ is the wavelength, $d^{(n)}_{k,m}$ is the distance of travel (DOT) for the corresponding link, and $\psi_{k}\sim\mathcal{U}(0,2\pi)$ accounts for the imperfect synchronization of the $k$-th UE, relative to the APs. Also,  $\rho(d)=\frac{\lambda_{o}}{4\pi d}$ and $\phi(d)=-\frac{2\pi}{\lambda_{o}}d$ correspond to the link attenuation and propagation phase, respectively. Then, the channel between the $k$-th UE and the $m$-th AP is 
\begin{align}\label{eq:h_km}
\mathbf{h}_{k,m} = [h^{(1)}_{k,m} , \ldots , h^{(n)}_{k,m}, \ldots , h^{(N_{\rm{AP}})}_{k,m}]\transp \in \mathbb{C}^{N_{\rm{AP}}\times 1}.
\end{align}

\subsection{System Architecture}
\vspace{-2mm}
Here, we introduce different system architectures, regarding the service allocation of the UEs and the distribution of the available antennas into several APs.
\subsubsection{Partially connected cell-free}
In partially connected cell-free systems, each AP serves a subset of the UEs ,which is a scalable architecture, since each AP has limited signaling overhead, in terms of CSI, precoding information etc. Let the sets $\mathcal{M}_{k}$ and $\mathcal{K}_{m}$ contain all the APs that serve the $k$-th UE and all the UEs served by the $m$-th AP, respectively. Consider the set of channels, as defined in~\eqref{eq:h_km}, for all APs in $\mathcal{M}_{k}$, as illustrated in Fig.~\ref{fig:sm}, which form a V-ELAA, placing the $k$-th UE in its near-field. This near-field condition enables better spatial resolution, allowing multiple UEs to be more effectively separated. Regarding the service allocation, two design strategies are proposed below:
\begin{enumerate}[label=\roman*., leftmargin=0.5cm]
\item \textbf{AP-centric CF (AP-C CF)}: Each AP serves its $K_{\rm{n}}$ closest UEs. However, this may leave some UEs without service.
\item \textbf{UE-centric CF (UE-C CF)}: Each UE selects its $M_{\rm{n}}$ closest APs. This can imbalance AP loads, overloading some APs surrounded by many UEs.
\end{enumerate}
Service allocation is based on statistical CSI or estimated UE positions and impacts both performance and complexity.  
\subsubsection{Fully Connected cell-free}
In fully connected cell-free (FC CF) all the APs serve all the UEs, simplifying the scheduling process. However, it is not a scalable solution. Essentially, the CF FC is an implementation of UC CF with $\mathcal{M}_{k}:\{1,\ldots,M\},\; \forall k$ and $\mathcal{K}_{m}:\{1,\ldots,K\},\; \forall m$. To simplify analysis, we consider a bound system that does not need to be scalable and the data exchange procedure among APs and the CPU is omitted.
\subsubsection{Cellular m-MIMO}
Centralizing all APs into a base station (BS) forms a single m-MIMO cell serving the entire area. The BS, with a UPA of $N_{\rm{BS}}$ antennas, has channels to each UE defined similarly to~\eqref{eq:h_km}. To simplify notation, the FC CF architecture directly extends to cellular m-MIMO by setting $M=1$ and $MN_{\rm{AP}} = N_{\rm{BS}}$.

\section{Channel Estimation \& Uplink communication}
Here, we examine \emph{structured} and \emph{unstructured} channel estimation, as well as \emph{decentralized} and \emph{centralized} combining.  

\subsection{Channel Estimation}
Consider a training procedure spanning over $T_{\rm{p}}$ channel uses with $T_{\rm{p}}$ available mutually-orthogonal pilot sequences; let $\mathbf{p}_{n}\!\in\!\mathbb{C}^{T_{\rm{p}}}$ be the $n$-th pilot sequence, normalized such that $\|\mathbf{p}_{n}\|^{2} = T_{\rm{p}}, \forall n$. We denote the index of the pilot assigned to UE $k$ as $n_k\!\in\!\{1,\ldots,T_{\rm{p}}\}$, and assume that $T_{\rm{p}}\!\geq\! K$, allowing orthogonal sequences with no pilot reuse among UEs, i.e., $\mathbf{p}_{n_{j}}\transp\mathbf{p}_{n_{k}}^{\ast} = 0$, for $n_j\neq n_k$. Then, the signal received by the $m$-th AP over the $T_{\rm{p}}$ channel uses, is organized in matrix form
\vspace{-2mm}
\begin{align}\label{eq:Y_m^p}
\mathbf{Y}_{m} = \sum_{k=1}^{K} \sqrt{P_{\rm{u}}}\mathbf{h}_{k,m}\mathbf{p}_{n_{k}}\transp + \mathbf{Z}_{m} \in\mathbb{C}^{N_{\rm{AP}}\times T_{\rm{p}}},
\end{align}
where $P_{\rm{u}}$ is the average UE transmit power per channel use and $\mathbf{Z}_{m} = [\mathbf{z}_{m}(1),\ldots,\mathbf{z}_{m}(t),\ldots,\mathbf{z}_{m}(T_{\rm{p}})]$ represents the noise, with $\mathbf{z}_{m}(t)\sim\mathcal{CN}(\mathbf{0},\sigma_{\rm{u}}^{2}\mathbf{I}_{N_{\rm{AP}}}), \forall t$. Post-multiplying~\eqref{eq:Y_m^p} by the conjugate pilot of the $k$-th UE yields
\vspace{-2mm}
\begin{align}\label{eq:y_tilde_km}
\tilde{\mathbf{y}}_{k,m} =& \mathbf{Y}_{m}\mathbf{p}_{n_{k}}^{\ast} 
= T_{\rm{p}}\sqrt{P_{\rm{u}}}\mathbf{h}_{k,m} + \tilde{\mathbf{z}}_{k,m}\in\mathbb{C}^{N_{\rm{AP}}\times 1},
\end{align}
exploiting the orthogonality among pilot sequences. Here, $\tilde{\mathbf{z}}_{k,m} = \mathbf{Z}_{m}\mathbf{p}_{n_{k}}^{\ast}$ is  a circularly-symmetric Gaussian random vector with covariance matrix $\mathbf{C}_{\rm{p}} = T_{\rm{p}}\sigma_{\rm{u}}^{2}\mathbf{I}_{N_{\rm{AP}}}$. The collected signals can be utilized to extract CSI, as described below.
\subsubsection{Unstructured Channel Estimation (UCE)}
Assuming no known channel structure, the CSI can be estimated individually for each UE-AP pair, utilizing~\eqref{eq:y_tilde_km}. The minimum square error channel estimation for the $k$-th UE is given by
\vspace{-2mm}
\begin{align}\label{eq:h_hat}
\hat{\mathbf{h}}_{k,m} =\frac{1}{T_{\rm{p}}\sqrt{P_{\rm{u}}}}\tilde{\mathbf{y}}_{k,m} = \mathbf{h}_{k,m} + \hat{\mathbf{z}}_{k,m}\in\mathbb{C}^{N_{\rm{AP}}\times 1}, 
\end{align}
where the estimation noise, $\hat{\mathbf{z}}_{k,m} = \tilde{\mathbf{z}}_{k,m}/(T_{\rm{p}}\sqrt{P_{\rm{u}}})$, is a circularly-symmetric Gaussian random vector with covariance matrix $\mathbf{C}_{\rm{u}} = \sigma_{\rm{u}}^{2}/(T_{\rm{p}}P_{\rm{u}})\mathbf{I}_{N_{\rm{AP}}}$.

\begin{algorithm}[H]
\caption{ML SCE for $k$-th UE and $m$-th AP.}
\begin{algorithmic}[1]\label{alg:str_csi_AP}
\REQUIRE $T_{\rm{p}},P_{\rm{u}}$ and $\tilde{\mathbf{y}}_{k,m}$ in~\eqref{eq:y_tilde_km}
\REPEAT
    \STATE Set: $\hat{\bm{\theta}} = \argmax\limits_{\bm{\theta}} \|\tilde{\mathbf{y}}_{k,m}\herm\bm{\alpha}(\hat{\bm{\theta}})\|^2$
    \STATE Set: $\hat{d} = \argmax\limits_{d} {\left(\|\tilde{\mathbf{y}}_{k,m}\|^2 - |N_{\rm{AP}}T_{\rm{p}}\sqrt{P_{\rm{u}}}\rho(d)|^{2}\right)^{-1}}$
    \STATE Set: $\hat{\psi} = \argmax\limits_{\psi} {\|\tilde{\mathbf{y}}-T_{\rm{p}}\sqrt{P_{\rm{u}}}\bm{h}_{\rm{L}}(\hat{d},\hat{\bm{\theta}},\psi)\|^{-2}}$
    \STATE Set: $\hat{d} = \argmax\limits_{d} {\|\tilde{\mathbf{y}}-T_{\rm{p}}\sqrt{P_{\rm{u}}}\bm{h}_{\rm{L}}(d,\hat{\bm{\theta}},\hat{\psi})\|^{-2}}$
\UNTIL{Convergence}
\RETURN  $\hat{\mathbf{h}}_{k,m} = \bm{h}_{\rm{L}}(\hat{d},\hat{\bm{\theta}},\hat{\psi})$
\end{algorithmic}
\end{algorithm}
\vspace{-3mm}

\subsubsection{Structured Channel Estimation (SCE)}
Exploiting the LoS structure, we propose a procedure, where each AP estimates its local channel. For convenience, we adopt the far-field assumption and the local channel model in~\eqref{eq:h_km} is represented as a function $\bm{h}_{\rm{L}}: (d,\bm{\theta},\psi)\mapsto\mathbf{h}$, for a DOT, $d$, an angle of departure (AOD), $\bm{\theta}=[\theta^{\rm{az}},\;\theta^{\rm{el}}]\transp$ along the azimuth and elevation, respectively, and a synchronization phase offset, $\psi$. Note that despite the far-field structure of each local channel, the collection of all APs construct a global channel for a V-ELAA that preserves near-field characteristics, resembling the sub-array processing of ELAAs~\cite{10541333}. The geometric parameters, $d,\bm{\theta}$ correspond to the reference element of the AP, $n^{\star}$. Then, the local channel at each AP is
\begin{equation}\label{eq:h_ff}
\bm{h}_{\rm{L}}(d,\bm{\theta},\psi) = \rho(d) \e^{\j\phi(d)} \e^{\j\psi} \bm{\alpha}_{\rm{f}}(\bm{\theta}) ,
\end{equation}
where $\bm{\alpha}_{\rm{f}}(\bm{\theta})\in\mathbb{C}^{N_{\rm{AP}}\times 1}$ is the far-field steering vector for $\bm{\theta}$.
To proceed, let us reformulate the observation in~\eqref{eq:y_tilde_km} as
\begin{align}\label{eq:y_tilde_ff}
\tilde{\mathbf{y}}_{k,m} = T_{\rm{p}}\sqrt{P_{\rm{u}}}\bm{h}_{\rm{L}}\left(d^{(n^{\star})}_{k,m},\bm{\theta}^{(n^{\star})}_{k,m},\psi_{k}\right) + \tilde{\mathbf{z}}_{k,m},
\end{align}
making explicit the dependence upon the channel parameters $d^{(n^{\star})}_{k,m}$, $\bm{\theta}^{(n^{\star})}_{k,m}$ and $\psi_{k}$. To simplify notation, let us abandon the index notation for the remainder of this discussion. For any post-processed signal $\tilde{\mathbf{y}}$ in~\eqref{eq:y_tilde_ff} we have a random vector with distribution $\tilde{\mathbf{y}}\sim\mathcal{CN}\left(T_{\rm{p}}\sqrt{P_{\rm{u}}}\bm{h}_{\rm{L}}\left(d,\bm{\theta},\psi\right),\mathbf{C}_{\rm{p}}\right)$. Thus, the likelihood function is given by
\begin{multline}\label{eq:likelihood}
\mathcal{L}_{\rm{L}}\!(\tilde{\mathbf{y}};\!d,\bm{\theta},\psi) \!=\!\!\frac{\exp\!\left\{\!\!-\!\left\|\mathbf{C}_{\rm{p}}^{\!-\!\frac{1}{2}}\!\Big(\!\tilde{\mathbf{y}}\!-\!T_{\rm{p}}\sqrt{\!P_{\!\rm{u}}}\bm{h}_{\rm{L}}\!\left(\!d,\bm{\theta},\psi\!\right)\!\!\Big)\!\right\|^2\!\right\}}{\pi^{N_{\rm{AP}}}\text{det}\{\mathbf{C}_{\rm{p}}\}},\!\!\!\!
\end{multline}
where $\text{det}\{\mathbf{C}_{\rm{p}}\} = (T_{\rm{p}}\sigma_{\rm{u}})^{N_{\rm{AP}}}$ and $\mathbf{C}_{\rm{p}}^{-\frac{1}{2}} = \sqrt{T_{\rm{p}}}\sigma_{\rm{u}}\mathbf{I}_{N_{\rm{AP}}}$. For any $\tilde{\mathbf{y}}$, the maximum-likelihood (ML) estimation of the corresponding channel is $\hat{\mathbf{h}} = \bm{h}_{\rm{L}}(\hat{d},\hat{\bm{\theta}},\hat{\psi}; \tilde{\mathbf{y}})$ , where
\begin{align}
(\hat{d},\hat{\bm{\theta}},\hat{\psi})&=\argmax_{d,\bm{\theta},\psi}\ \mathcal{L}_{\rm{L}}(\tilde{\mathbf{y}};d,\bm{\theta},\psi)\notag\\
&= \argmin_{d,\bm{\theta},\psi}\ \left\|\tilde{\mathbf{y}}-T_{\rm{p}}\sqrt{P_{\rm{u}}}\bm{h}_{\rm{L}}\left(d,\bm{\theta},\psi\right)\right\|^2.\label{eq:decent_ML}
\end{align}
The described optimization problem can be solved through alternating optimization, exploiting the decoupled structure of $\bm{h}_{\rm{L}}(d,\bm{\theta},\psi)$ in the angular and range domain. The procedure is summarized in Alg.~\ref{alg:str_csi_AP}. The optimization problem in~\eqref{eq:decent_ML} is non-convex and the computationally friendly Alg.~\ref{alg:str_csi_AP} does not necessarily converge to the global maximum.

\subsection{Uplink Communication}
\vspace{-1mm}
Now, let us consider an uplink communication procedure that spans $T_{\rm{u}}$ intervals. Then, the received signal is
\begin{align}\label{eq:y_m_ul}
\mathbf{y}_{m}(t) =& \sqrt{P_{\rm{u}}}\sum_{k=1}^{K} \mathbf{h}_{k,m}x_{k}(t) + \mathbf{z}_{m}(t),
\end{align}
during the $t$-th interval at the $m$-th AP, where $x_{k}(t)$ is the uplink data symbol and the noise $\mathbf{z}_{m}(t)$ is a circularly-symmetric Gaussian random vector with covariance $\sigma_{\rm{u}}^{2}\mathbf{I}_{N_{\rm{AP}}}$. 

In a \emph{decentralized} network, the $m$-th AP processes~\eqref{eq:y_m_ul} with a locally constructed combining vector $\mathbf{v}_{k,m}$ for the $k$-th UE and forwards the estimated data for decoding. We also define the auxiliary matrix $\mathbf{A}_{k,m} = \mathbf{I}_{N_{\rm{AP}}}$, if $m\in\mathcal{M}_{k}$ and $\mathbf{A}_{k,m} = \mathbf{0}_{N_{\rm{AP}}}$, otherwise.
Also, let $\bm{\eta}_{k} = [\eta_{k,1}, \ldots, \eta_{k,M}]\transp\in\mathbb{C}^{M\times1}$, where $\eta_{k,m}$ is a scalar weight for the corresponding local estimate. The estimated symbol of the $k$-th UE is 
\vspace{-1mm}
\begin{align}
\hat{x}_{k}(t) &= \sum\nolimits_{m=1}^{M}\eta_{k,m}^{\ast} \mathbf{v}_{k,m}\herm\mathbf{A}_{k,m}\mathbf{y}_{m}(t),   \\
&= \sqrt{P_{\rm{u}}}\bm{\eta}_{k}\herm\bm{\zeta}_{k,k}x_{k}(t) + \sqrt{P_{\rm{u}}}\sum\limits_{j\neq k}^{K}\bm{\eta}_{k}\herm\bm{\zeta}_{k,j}x_{j}(t) + \mathbf{z}_{k}'(t), \notag
\end{align}
where $\bm{\zeta}_{k,j} = [\mathbf{v}_{k,1}\herm\mathbf{A}_{k,1}\mathbf{h}_{j,1}, \ldots, \mathbf{v}_{k,M}\herm\mathbf{A}_{k,M}\mathbf{h}_{j,M}]\transp\in\mathbb{C}^{M\times 1}$ and $\mathbf{z}_{k}'(t) = \sum_{m=1}^{M}\eta_{k,m}^{\ast} \mathbf{v}_{k,m}\herm\mathbf{A}_{k,m}\mathbf{z}_{m}(t)$. Then, for the local \emph{decentralized} procedure, an achievable uplink spectral efficiency (SE) for the $k$-th UE is given by
\vspace{-1mm}
\begin{align}\label{eq:se_k_u-D}
\text{SE}_{k}^{\rm{u-D}}\ = \frac{T_{\rm{u}}}{T_{\rm{c}}}\log_{2}\left(1 + \gamma_{k}^{\rm{u-D}}\right),
\end{align}
where $\gamma_{k}^{\rm{u-D}}$ is the effective signal-to-interference-plus-noise ratio (SINR) for the described system architecture, given by
\begin{align}\label{eq:gamma_k_u-D}
\gamma_{k}^{\rm{u-D}}=\frac{\left|\bm{\eta}_{k}\mathbb{E}\left\{ \bm{\zeta}_{k,k} \right\}\right|^2}{\bm{\eta}_{k}\herm \Xi_{k}\bm{\eta}_{k}},
\end{align}
where we define $\bar{\mathbf{v}}_{k,m} = \mathbf{A}_{k,m}\mathbf{v}_{k,m}$ and 
\vspace{-2mm}
\begin{align}\label{eq:exp_Zeta_kj}
\Xi_{k} = \sum_{j=1}^{K}\mathbb{E}\left\{ \bm{\zeta}_{k,j}\bm{\zeta}_{k,j}\herm\right\} - \mathbb{E}\left\{\bm{\zeta}_{k,k}\right\}\mathbb{E}\left\{\bm{\zeta}_{k,k}\herm\right\} + \mathbf{F}_{k},
\end{align}
with $\mathbf{F}_{k} = \frac{\sigma_{\rm{u}}^{2}}{P_{\rm{u}}}\diag\left(\mathbb{E}\{\|\bar{\mathbf{v}}_{k,1}\|^{2},\ldots,\mathbb{E}\{\|\bar{\mathbf{v}}_{k,M}\|^{2}\right)$.

For a \emph{centralized} network, the CPU combines~\eqref{eq:y_m_ul} from all APs, $\mathbf{y}(t) = [\mathbf{y}_{1}(t)\transp \ldots \mathbf{y}_{M}(t)\transp]\transp\in\mathbb{C}^{MN_{\rm{AP}}\times 1}$, before decoding the data for all UEs. Let $\mathbf{v}_{k}\in\mathbb{C}^{MN_{\rm{AP}}\times 1}$ be the $k$-th combining vector and $\mathbf{A}_{k} = \diag(\mathbf{A}_{k,1},\ldots,\mathbf{A}_{k,M})$. Then,
\vspace{-2mm}
\begin{align}
\hat{x}_{k}(t) =& \mathbf{v}_{k}\herm\mathbf{A}_{k}\mathbf{y}(t),
\end{align}
is the estimated symbol of the $k$-th UE and the uplink SE is
\begin{align}\label{eq:se_k_u-C}
\text{SE}_{k}^{\rm{u-C}}\ = \frac{T_{\rm{u}}}{T_{\rm{c}}}\log_{2}\left(1 + \gamma_{k}^{\rm{u-C}}\right),
\end{align}
where $\gamma_{k}^{\rm{u-C}}$ is the effective SINR, given by
\begin{align}\label{eq:gamma_k_u-C}
\gamma_{k}^{\rm{u-C}}\!\!=\frac{\left|\mathbb{E}\left\{ \xi_{k,k} \right\}\right|^2}{\sum\limits_{j=1}^{K}\mathbb{E}\left\{ \left| \xi_{k,j} \right|^2\right\} - \left|\mathbb{E}\left\{ \xi_{k,k} \right\}\right|^2 + \frac{\sigma_{\rm{u}}^{2}}{P_{\rm{u}}}\mathbb{E}\left\{ \|\bar{\mathbf{v}}_{k}\|^2 \right\}}.    
\end{align}
Note that $\xi_{k,j} = \mathbf{v}_{k}\mathbf{A}_{k}\mathbf{h}_{j}$ and $\bar{\mathbf{v}}_{k} = \mathbf{A}_{k}\mathbf{v}_{k}$ and the statistical expectation in~\eqref{eq:gamma_k_u-C} is taken with respect to the constructed combining vectors under imperfect CSI.


\section{Numerical Analysis}
We evaluate the performance of the considered network architectures, as shown in Fig.~\ref{fig:sm}, under perfect and imperfect CSI conditions. The cumulative distribution functions (CDFs) are produced using Monte Carlo simulations with multiple random AP and UE placements within a $500$ m$\times500$ m area. The UEs' heights are between 1-2 m, while APs are placed at 12-13 m with a random array orientation, $\varphi_{m}\sim\mathcal{U}(0,2\pi)$. The number of UEs varies as $K = \{20,40,60,80\}$, with AP configurations of $M = \{1, 4, 25\}$, where each AP has $N_{\rm{AP}} = \{100,25,4\}$ antennas, ensuring a constant total number of antennas across scenarios. The system operates at $f_{o}=30$ GHz carrier frequency ($\lambda_{o}=1$ cm) with $B=20$ MHz bandwidth, 13 dBm uplink transmit power, and a 7 dB noise figure. Training and uplink intervals are set to $T_{\rm{p}} = K$ and $T_{\rm{u}} = 100$, respectively. The antenna spacing is $\delta_{o} = \lambda_{o}/2$ and the first element acts as reference, i.e., $n^{\star}=1$. Even under the wideband assumption, phase variations around the carrier cause negligible changes for each LoS link.

\subsection{Scalable Combining Schemes}
Before we proceed, let us define commonly used designing strategies for the combining vectors, including maximum rate (MR) and partial regularized zero-forcing (P-RZF). For the decentralized combining procedure, the MR combining vector dedicated to the $k$-th UE, at the $m$-th AP is structured as
\begin{align}\label{eq:v_k_mr_D}
\mathbf{v}_{k,m}^{\rm{mr}} = \mathbf{A}_{k,m}\hat{\mathbf{h}}_{k,m},
\end{align}
which maximizes $\mathbb{E}\left\{ \bm{\zeta}_{k,k} \right\}$. Similarly, for the centralized procedure the $k$-th MR combining vector at the CPU is 
\begin{align}\label{eq:v_k_mr_C}
\mathbf{v}_{k}^{\rm{mr}} = \mathbf{A}_{k}\hat{\mathbf{h}}_{k},
\end{align}
which maximizes the numerator of the SINR expression in~\eqref{eq:gamma_k_u-C}. However, the inter user interference (IUI) is not addressed. A more practical solution that takes IUI into account is the P-RZF. For the decentralized procedure, the $k$-th P-RZF combining vector, at the $m$-th AP,  
\begin{align}\label{eq:v_k_przf_D}
\mathbf{v}_{k,m}^{\rm{zf}}&= P_{\rm{u}}\biggl(\sum_{j\in\mathcal{K}_{m}}P_{\rm{u}}\mathbf{A}_{k,m}\hat{\mathbf{h}}_{j,m}\hat{\mathbf{h}}_{j,m}\herm\mathbf{A}_{k,m} +\sigma_{\rm{u}}^{2}\mathbf{I}_{N_{\rm{AP}}}\biggr)^{\!\!-1}\notag\\
&\ \ \ \times\mathbf{A}_{k,m}\hat{\mathbf{h}}_{k,m}.
\end{align}
The combining weight vector $\bm{\eta}_{k}$ can be designed by the CPU to maximize the SINR expression in~\eqref{eq:gamma_k_u-D}, based on channel statistics. The optimal weight vector for the $k$-th UE is 
\begin{align}\label{eq:eta_k_star}
\bm{\eta}_{k}^{\star} = P_{\rm{u}}\bigg(\sum_{j=1}^{K}\mathbb{E}\left\{\bm{\zeta}_{k,j}\bm{\zeta}_{k,j}\herm\right\} + \mathbf{F}_{k} + \bar{\mathbf{A}}_{k} \bigg)^{-1}\mathbb{E}\left\{\bm{\zeta}_{k,k}\right\},
\end{align} 
where $\bar{\mathbf{A}}_{k}\in\mathbb{R}^{M\times M}$ is a modified identity matrix, such that the $(m,m)$-th element is zero if $m\in\mathcal{M}_{k}$. Utilizing $\bm{\eta}_{k}^{\star}$ at the CPU, the SINR expression in ~\eqref{eq:gamma_k_u-D} is simplified to 
\begin{align}\label{eq:gamma_k_u-D_star}
\gamma_{k}^{\rm{u-D}\star}= \mathbb{E}\left\{\bm{\zeta}_{k,k}\herm\right\} \big(\Xi_{k} + \frac{1}{P_{\rm{u}}}\bar{\mathbf{A}}_{k} \big)^{-1} \mathbb{E}\left\{\bm{\zeta}_{k,k}\right\}.
\end{align}   
 
For the centralized procedure, the $k$-th P-RZF combining vector at the CPU is 
\begin{align}\label{eq:v_k_przf_C}
\mathbf{v}_{k}^{\rm{zf}}  = P_{\rm{u}}\bigg( \sum_{j\in\mathcal{Z}_{k}}P_{\rm{u}}\mathbf{A}_{k}\hat{\mathbf{h}}_{j}\hat{\mathbf{h}}_{j}\herm\mathbf{A}_{k} + \sigma_{\rm{u}}^{2}\mathbf{I}_{MN_{\rm{AP}}}\bigg)^{-1}\!\!\!\!\!\mathbf{A}_{k}\hat{\mathbf{h}}_{k},
\end{align}
where the set $\mathcal{Z}_{k}=\{j:\trace(\mathbf{A}_{k}\mathbf{A}_{j})\neq0\}$ contains the UEs that are \emph{strong sources of IUI} for the $k$-th UE. 

\subsection{Numerical results}
\begin{figure}[t]
    \centering
    \includegraphics[width=0.90\columnwidth]{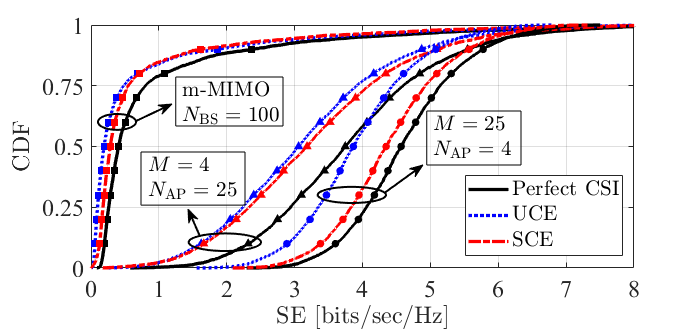}
    \caption{CDF for SE utilizing P-RZF, for different channel estimation strategies and $K=40$. The number of APs increases, as indicated by the arrows, while the total number of antennas remains constant.}
    \label{fig:se_ul_przf_c_size}
    \vspace{-0.25cm}
\end{figure}
Here, we present numerical results for various simulation setups. Fig.~\ref{fig:se_ul_przf_c_size} depicts the CDF of the achievable SE in the FC CF architecture for different AP configurations. We compare a single m-MIMO BS setup (\scalebox{0.75}{$\blacksquare$}: $M=1, N_{\rm{BS}}=100$) with medium / sparse (\scalebox{1.25}{$\bullet$}: $M=25, N_{\rm{AP}}=4$) and small / dense ($\blacktriangle$: $M=4, N_{\rm{AP}}=25$) AP deployments, keeping the total antenna count constant and setting $K=40$ UEs. The results show that denser AP deployments enhance SE across all CSI scenarios: perfect CSI (solid black curves), SCE (red dash-dot curves), and UCE (blue dotted curves). As expected, perfect CSI achieves the highest SE, while SCE consistently outperforms UCE, particularly in dense deployments.  

\begin{figure}[t]
    \centering
    \includegraphics[width=0.90\columnwidth]{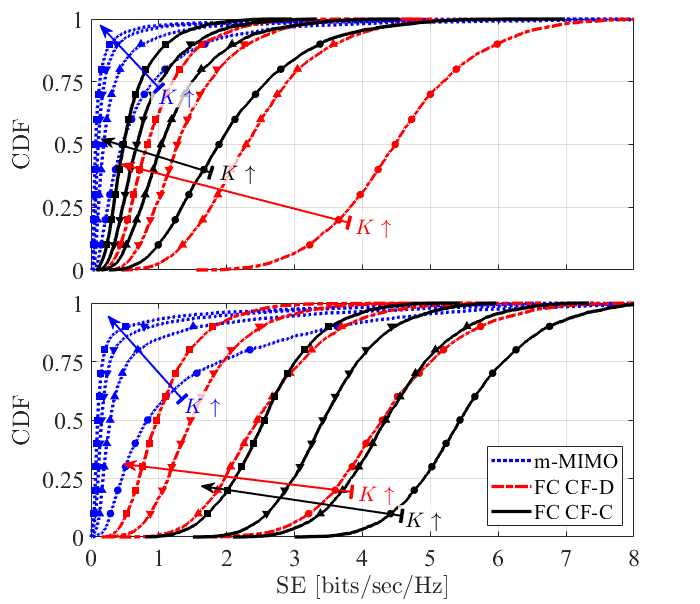}
    \caption{CDF for SE utilizing MR (top) and P-RZF (bottom), for different system architectures. The number of UEs increases, as indicated by the arrows, with $K=\{20,40,60,80\}$ corresponding to markers \{\scalebox{1.25}{$\bullet$}, $\blacktriangle$, $\blacktriangledown$, \scalebox{0.75}{$\blacksquare$}\}.}
    \label{fig:se_ul_mr_przf_K_up}
    \vspace{-0.25cm}
\end{figure}

Fig.~\ref{fig:se_ul_mr_przf_K_up} shows the achievable SE under MR and P-RZF combining with imperfect CSI and SCE for $M=25$ APs and FC CF. We compare centralized (FC CF-C, solid black curves) and decentralized (FC CF-D, red dash-dot curves) architectures, alongside the m-MIMO case (blue dotted curves). The number of UEs increases, as indicated by the arrows, with $K=\{20,40,60,80\}$ corresponding to markers \{\scalebox{1.25}{$\bullet$}, $\blacktriangle$, $\blacktriangledown$, \scalebox{0.75}{$\blacksquare$}\}.

With MR combining (Fig.~\ref{fig:se_ul_mr_przf_K_up}-top), the FC CF-D architecture consistently outperforms FC CF-C, due to the combining weight vector, $\bm{\eta}_{k}$, which prioritizes UEs with strong channels at each AP. In contrast, FC CF-C attempts to combine all data directly. The m-MIMO case performs the worst due to significant IUI, even for a small number of UEs.

Conversely, with P-RZF combining (Fig.~\ref{fig:se_ul_mr_przf_K_up}-bottom), FC CF-C architecture achieves the best SE, effectively mitigating IUI even with a large number of UEs. FC CF-D and m-MIMO cases also improve compared to their MR-based counterparts. However, m-MIMO remains unable to provide adequate SE for most UEs as their number increases.

\begin{figure}[t]
    \centering
    \includegraphics[width=0.90\columnwidth]{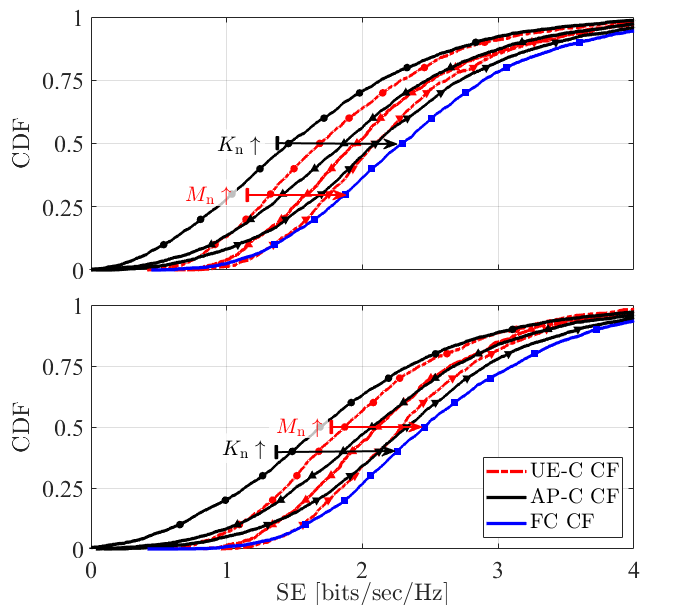}
    \caption{CDF for SE utilizing MR (top) and P-RZF (bottom) for PC CF architectures for $K=40$. The number of served UEs per AP ($K_{\rm{n}}$) in AP-C CF and the number of connected APs per UE ($M_{\rm{n}}$) in UE-C CF increases, as indicated by the arrows, with $K_{\rm{n}}=\{10,20,30\}$ and $M_{\rm{n}}=\{6,13,19\}$ corresponding to markers \{\scalebox{1.25}{$\bullet$}, $\blacktriangle$, $\blacktriangledown$\}. The FC CF system is marked in blue.}
    \label{fig:se_ul_mr_przf_ap_ue}
    \vspace{-0.25cm}
\end{figure}

Finally, Fig.~\ref{fig:se_ul_mr_przf_ap_ue} presents the achievable SE for $K=40$ UEs and $M=25$ APs under decentralized MR (top) and P-RZF (bottom) combining. The comparison between AP-C CF (solid black curves) and UC-C CF (red dash-dot curves) is conducted under imperfect CSI and SCE conditions. The SE is depicted as a function of the number of UEs served per AP ($K_{\rm{n}}$) and the number of APs serving each UE ($M_{\rm{n}}$), as indicated by the arrows. Notably, UC-C CF is plotted against the right y-axis and the top x-axis, which is reversed.  

As $K_{\rm{n}}$ and $M_{\rm{n}}$ increase, the SE consistently improves for both MR and P-RZF combining schemes. A comparison between the subplots in Fig.~\ref{fig:se_ul_mr_przf_ap_ue} reveals that the two combining strategies yield similar performance in a decentralized partially connected CF system. Interestingly, when $K_{\rm{n}}$ and $M_{\rm{n}}$ are large, the SE approaches that of FC CF ($K_{\rm{n}}=K$, $M_{\rm{n}}=M$), demonstrating that a scalable CF architecture enables a flexible and efficient system design. This result highlights that even with simple combining strategies, the system can maintain excellent performance while requiring minimal signaling overhead between APs and the CPU.

\section{Conclusions}
In this study, we analyzed various system architectures, including PC CF, FC CF, and m-MIMO operating at mmWave. We examined the potential for near-field capabilities in CF LoS scenarios, where APs form V-ELAAs capable of resolving and serving numerous UEs. Our investigation focused on the structure of the channel and the degrees of freedom it offers for channel estimation, highlighting the superiority of SCE over conventional UCE schemes. Additionally, we evaluated decentralized and centralized combining schemes, assessing their performance under MR and P-RZF combining strategies. Finally, in the PC CF system, we explored both AP- and UE-centric scheduling strategies, analyzing their performance as they converge toward FC CF systems. The presented LoS-based analysis validates CF performance limits relative to m-MIMO. Future work should consider more general channel models, including Non-LoS links.

\bibliographystyle{ieeetr}
\bibliography{references}

\end{document}